\documentclass[twocolumn, 
twocolappendix]{aastex701}

\usepackage[utf8]{inputenc}
\usepackage[caption=false]{subfig}
\usepackage{amsmath}
\usepackage{placeins}

\newcommand{\mjb}{\,mJy\,beam$^{-1}$}
\newcommand{\ujb}{\,$\mu$Jy\,beam$^{-1}$}

\newcommand{\cxo}{\emph{Chandra}}

\newcommand{\psr}{PSR~J2229+6114}

\begin{document}

\title{High-resolution Very Large Array Radio Observations of the Boomerang Pulsar Wind Nebula}

\author[0000-0003-3601-5127]{Paul C. W. Lai}
\email{chong.lai.22@ucl.ac.uk}
\affiliation{Department of Physics, The University of Hong Kong, Pokfulam Road, Hong Kong}
\affiliation{Mullard Space Science Laboratory, University College London, 
Holmbury St Mary, Surrey RH5 6NT, United Kingdom}

\author[0000-0002-5847-2612]{C.-Y. Ng}
\affiliation{Department of Physics, The University of Hong Kong, Pokfulam Road, Hong Kong}
\affiliation{Hong Kong Institute for Astronomy and Astrophysics, The University of Hong Kong, Pokfulam Road, Hong Kong}
\email{ncy@astro.physics.hku.hk}

\author[0000-0002-2096-6051]{Shumeng Zhang}
\affiliation{Department of Physics, The University of Hong Kong, Pokfulam Road, Hong Kong}
\affiliation{Hong Kong Institute for Astronomy and Astrophysics, The University of Hong Kong, Pokfulam Road, Hong Kong}
\email{shumeng_zhang@connect.hku.hk}

\begin{abstract}
We present a radio polarimetric study of the Boomerang pulsar wind nebula
G106.65+2.96 with Very Large Array (VLA) observations at the 6\,GHz band. 
Our high-resolution image
discovers new small-scale features in the nebula, including an elliptical core
of $40''\times20''$ surrounding the central pulsar and a $2'$-long
arc wrapping around the core in the north. The latter shows a gap from the core, and it consists of a bright lobe in the northwest and a tongue-like structure
in the northeast. These could be resulting from the pulsar wind interaction with the
environment. Our polarization measurement reveals a highly ordered magnetic
field with toroidal geometry. The small scale features are all highly linearly
polarized. In particular, the lobe has a polarization fraction of $\sim$60\%,
close to the synchrotron limit. This is also much higher than the value measured
at a lower frequency, implying significant depolarization. We show that
this can be explained by Faraday rotation in the nebula, and we constructed a
simple 3D model accordingly to estimate a magnetic field strength of
$\sim$25--70\,$\mu$G assuming a median inclination angle of $60^\circ$, which uncertainty arises from the uncertainty in the flux measurement and its distance.
 
\end{abstract}


\section{Introduction} \label{sec:intro}
A pulsar is a rapidly rotating, highly magnetized neutron star produced in a supernova
explosion at the end stage of massive stellar evolution. As an analog to a
rotating magnet, a pulsar can create a huge electric potential around it. This
accelerates charged particles to relativistic speeds, driving a magnetized
outflow known as a pulsar wind \citep[see][for a review]{Amato2024PulsarAcceleration}. The
wind mainly composes of electrons and positrons. When these particles interact
with the surrounding medium, either the interstellar medium or the supernova
remnant (SNR), a termination shock is formed, and they are further accelerated to
ultra-relativistic energies at the shock. The high energy particles then inflate
a synchrotron bubble emitting from radio to X-ray bands.
Such a structure is called a pulsar wind nebula (PWN).

The fast rotation of a pulsar wraps its magnetic field lines, making them
nearly toroidal before reaching the termination shock \citep{conto+99}.
Downstream of the shock, the plasma flow becomes turbulent and a poloidal
$B$-field component emerges \citep{porth+14,olmi+16,Olmi23}. Observationally,
this can be tested with polarimetric measurements, since PWN emission is
dominated by synchrotron emission which is highly linearly polarized and the
polarization angle can reveal the local magnetic field direction at the
emission site. Previous observations of PWNe show a vast variety of magnetic
field configurations, including a radial field \citep[e.g., DA 495 and
G21.5$-$0.9;][]{kothes+08,lai+22}, a large-scale toroidal field \citep[e.g.,
Vela;][]{dlmd03}, an elongated field aligned with the pulsar spin axis
\citep[e.g., 3C58;][]{reich02}, and other complex structures \citep[e.g., the
Crab and G292.0+1.8;][]{crab_mag, gaensler+wallace03}. The diverse behavior is
very difficult to be explained by a unified picture.

The Boomerang (G106.6+2.9 or G106.65+2.96) is a middle-aged PWN located at the
tip of the associated SNR G106.3+2.7 \citep{kothes+01, Kothes2006A&A}. It is powered by pulsar
J2229+6114 which has spin period $P=51.6$\,ms, spin-down rate $\dot
P=7.83\times10^{-14}$, surface magnetic field $B=2.0\times10^{12}$\,G, spin-down
luminosity $\dot E=2.2\times10^{37}$\,erg\,s$^{-1}$, and characteristic age
$\tau_c=10.5$\,kyr \citep{halpern+01a}. In the radio band, the PWN shows an
arc-like overall morphology of $\sim3'$ long that resembles a boomerang
\citep{kothes06}. This shape is highly unusual among PWNe, as the emission is
found only on one side of pulsar, and the peak is significantly offset from the
pulsar location. This is believed to be resulting from crushing of the nebula by
the supernova reverse shock $3.9\,$kyr ago \citep{kothes06}. The X-ray image of the
Boomerang reveals a drastically different structure: 
the arc-minute-scale radio emission
is not detected, but there is an X-ray arc of $\sim20''$ with a jet-like
feature surrounding the pulsar \citep{halpern+01a}, which is interpreted as the
toroidal termination shock \citep{ng04,ng08}. Beyond that, there is very faint,
large-scale diffuse X-ray emission extending south \citep{ge+21}. Such
morphology indicates that the X-ray nebula could be a revived or reborn PWN
after the passage of the SNR reverse shock.

The Boomerang has drawn much attention in gamma-ray studies lately. Ultra-high energy photons up to 570\,TeV are detected in the direction of the
SNR, suggesting that this is a PeVatron---an astrophysical source that can
accelerate particles up to PeV energies \citep{boom_lhaaso_2021}. The exact
production site and mechanism of the gamma-rays are still under debate. Previous
radio observations discovered a dense H{\sc i} shell surrounding the Boomerang
\citep{kothes+01}, which provides some support for a hadronic origin
\citep{ge+21,yang+22}, but a leptonic origin from pulsar wind is also possible
\citep[see][]{liu+20}.

 
Single-dish radio measurements found a spectral break for the Boomerang around
4.3\,GHz \citep{kothes06}. 
If interpreted as the
synchrotron cooling break, the magnetic field strength of the Boomerang would be
2.6\,mG, much stronger than the typical values of 10--100\,$\mu$G
\citep[e.g.,][]{snail,Liu23ApJ}. Later studies argue for a much weaker field
from $3\,\mu$G to $140\,\mu$G based on pressure estimate \citep{Chevalier+11}
and modeling of the broadband spectrum from radio to gamma-rays \citep{liang+22,
Pope2024ApJ}. This implies that the 4.3\,GHz break could be intrinsic instead of
due to cooling. Radio polarization measurements found a toroidal configuration
for the projected $B$-field, similar to that of the Vela PWN \citep{dlmd03}.
However, the distribution of rotation measure (RM) in the nebula prefers a
radial field instead \citep{kothes06}. 

In this paper, we revisit the Boomerang PWN using high resolution radio
observations taken with the Karl G. Jansky Very Large Array (VLA) to further
investigate the nebular magnetic field structure. We note that in the literature
there are various distance estimates for this source, ranging from 0.8 to
12\,kpc \citep[see][and references therein]{Pope2024ApJ}. 
\cite{Pineault2000AJ} suggested that the SNR led to a deficit of neutral hydrogen at an LSR velocity about $-105$\,km\,s$^{-1}$, and estimated that the distance to the SNR was 12\,kpc.
Later studies used H{\sc i} maps with
    better resolutions and 
    found a deficit of neutral
    hydrogen at a much smaller distance
    $\sim0.8\,$pc \citep{kothes+01, kothes06}.
A distance of 12\,kpc puts the SNR in an extreme size ($\sim160$\,pc) and gamma-ray brightness \citep{Pope2024ApJ}, which is less favorable than a shorter distance.
\cite{Pope2024ApJ} suggest a distance of 8\,kpc, which appears most consistent with the PWN spectral energy distribution model.
We do not attempt to constrain the distance in this work and will present the uncertainty arises from distance measurements wherever needed.
This paper is organized as following: the
observation details and imaging method are described in Section~\ref{sec:data}
and the results are presented in Section~\ref{sec:obs_results}. In
Section~\ref{sec:discuss}, we present an interpretation of the PWN morphology,
and we model the magnetic field strength and structure based on the
observational results. Finally, we summarize our findings in
Section~\ref{sec:conclusion}.

\section{Observation and Data Reduction} \label{sec:data}

\begin{deluxetable*}{cccccc} 
\tablecaption{VLA observations used in this study}
 \label{tab:obs}
\tablehead{
\colhead{Observing date} & \colhead{2017 Nov 3} & \colhead{2017 Nov 12} &
\colhead{2017 Nov 21} & \colhead{2021 Jul 11}}
\startdata
VLA array configuration & B & B & B & C \\
Center frequency (GHz) & 6.0 & 6.0 & 6.0 & 6.0 \\
Bandwidth (GHz) & 4.0 & 4.0 & 4.0 & 4.0\\
On-source time (min) & 40 & 95 & 95 & 60 \\
Baseline (km) & 0.18--9.9 & 0.12--10.3 & 0.24--10.7 & 0.03--3.1 \\
$u$-$v$ coverage (k$\lambda$) & 2.5--260 & 2--270 & 3.5--280 & 0.42--82 \\
Flux calibrator & 3C138 & 3C286 & 3C138 & 3C138 \\
Phase calibrator & J2148+6107 & J2148+6107 & J2148+6107 & J2148+6107 \\
Polarization angle calibrator & 3C138 & 3C286 & 3C138 & 3C138 \\
Polarization leakage calibrator & J2355+4950 & J1407+2827 & J2355+4950 & J2355+4950
\enddata
\end{deluxetable*}

The Boomerang PWN was observed with the Karl G. Jansky Very Large Array
(hereafter: VLA) in the C band (4--8\,GHz). The observations were taken in the B
array configuration on 2017 Nov 3, 12, and 21, with 230\,min total on-source
time and in the C array configuration on 2021 Jul 11, with 60\,min on-source
time. We combined all data to produce high-resolution images. The $u$-$v$
coverage and calibrators of the observations are listed in Table~\ref{tab:obs}.
Given the shortest baseline is
$\sim0.03$\,km, in theory the largest angular scale that can be imaged is $\sim4\arcmin$.
However, this is only true for a full 12-hour synthesis observation.
Our C-array observation was one hour long and the largest angular scale it could probed should be halved\footnote{Based on the VLA observation manual: \url{https://science.nrao.edu/facilities/vla/docs/manuals/oss/performance/resolution}}, which is $\sim2\arcmin$.
This is smaller than the size of the Boomerang ($\lesssim3'$).
Therefore, the flux measurement of this observation will be underestimated.

All data reduction was done using the Common Astronomical Software Application
(CASA) version 6.2.1 \citep{casa}. The data were first processed with the
pipeline for preliminary flaggings and standard calibrations. We then further
examined and flagged the remaining suspected RFI signals, and re-calibrated the
data manually. 
The polarization calibration was also performed manually at this stage.

All the images were generated and cleaned using the CASA task \texttt{tclean}.
To optimize between sensitivity and resolution, we chose the Briggs weighting
algorithm with robust $=1.0$ for all the images. The Stokes I, Q, and U maps at
6\,GHz were produced by combining all the available data using the full
bandwidth. Since we mainly focus on the extended emission of the PWN, we applied
a Gaussian $uv$-taper of $80\,{\rm k}\lambda$ to effectively smooth the images
and increase sensitivity towards extended emission. The cleaned images are
restored and then corrected for primary beam sensitivity. The linear
polarization intensity (PI) map is obtained by combining the Stokes Q and U maps
(${\rm PI} = \sqrt{{\rm Q}^2 + {\rm U}^2}$), with the Ricean bias corrected
\citep{ricean}. Our final Stokes I and PI images at 6\,GHz have a beam size of
FWHM $2\farcs3\times1\farcs8$ and rms noise of 5\ujb. The latter is compatible
with the theoretical estimate.

Forming images using the entire 4\,GHz bandwidth improves the $u$-$v$ coverage,
but this could lead to slight bandwidth depolarization. Previous studies found
that the rotation measure (RM) of the nebula is around $-$200\,rad\,m$^{-2}$
\citep{kothes06}, therefore combining the 4--8\,GHz data would cause $\sim7\%$
depolarization\footnote{The bandwidth depolarization fraction is given by
$f_{\rm depol} = 1 - \sin{(\Delta \theta)}/\Delta \theta$, where $\Delta \theta
= 2{\rm RM}c^2 \Delta \nu/\nu_{\rm c}^3$, $\nu_{\rm c}$ is the center frequency
and $\Delta \nu$ is the bandwidth.}. We argue that this level is acceptable, as
it is comparable to the measurement uncertainties of our observations.

To correct for foreground Faraday rotation, we divided the full band into four
1\,GHz-wide subbands to generate the polarization position angle (PA) maps. The
same imaging and deconvolution procedures as above were employed, but we reduced the spatial resolution to boost the signal-to-noise ratio, by
employing a $uv$-taper of 40\,k$\lambda$ and a restoring beam of FWHM
$5\farcs0\times5\farcs0$. The observed PA of the polarization vectors is
calculated from the Stokes Q and U maps as ${\rm PA}=\frac{1}{2}\arctan{({\rm
U}/{\rm Q})}$. A linear fit to PA values with respect to the square of
wavelength ($\lambda^2$) gives the RM map and the intrinsic PA. We discard any
fits that have RM uncertainty over 45\,rad\,m$^{-2}$ or intrinsic PA uncertainty
over $15^\circ$, or if the total intensity is fainter than 15\ujb\ at 6\,GHz.

\section{Observation Results} \label{sec:obs_results}
\subsection{Source structure}
\begin{figure*}
 \centering
 \includegraphics[width=0.98\textwidth]{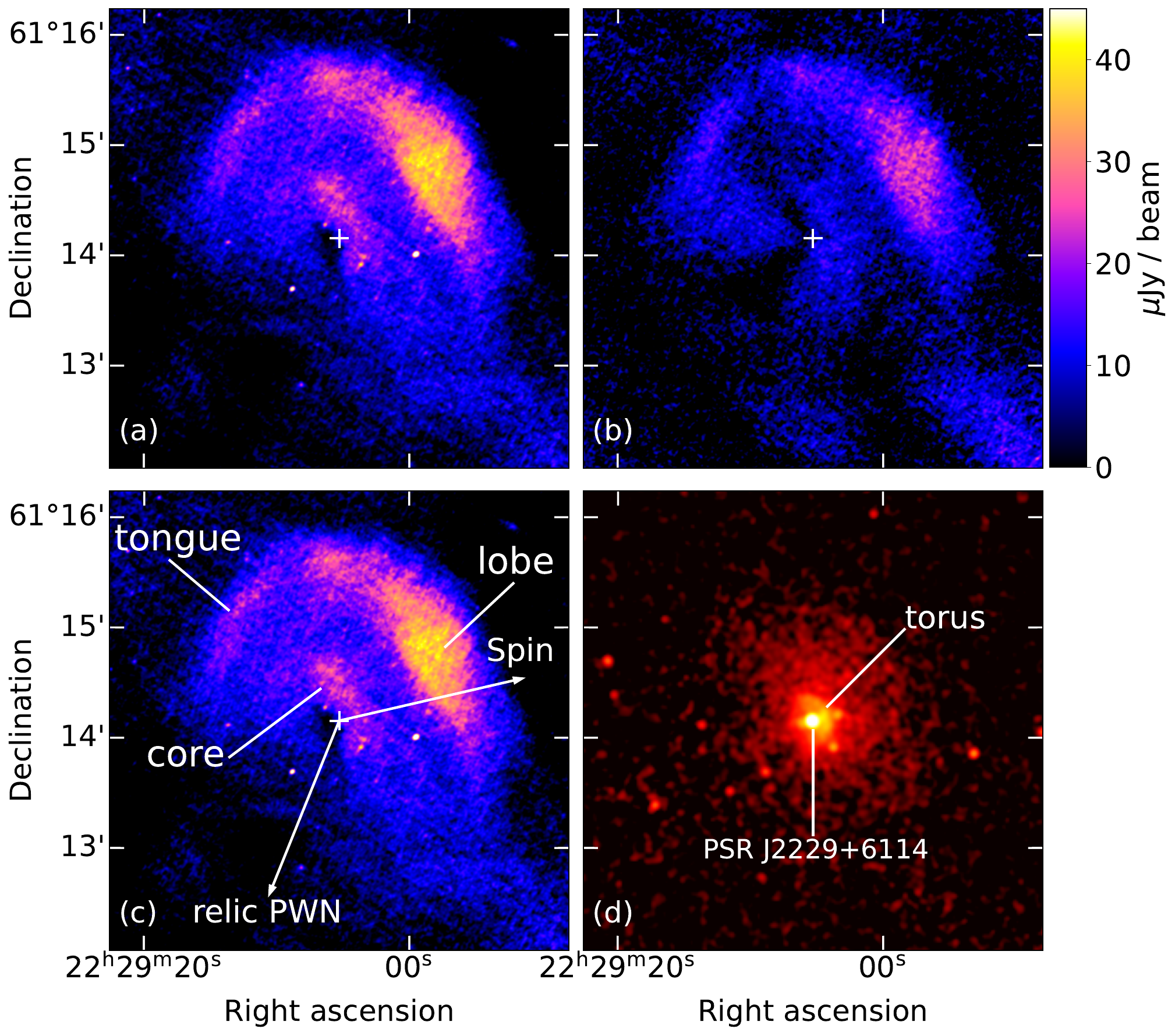}
 \caption{(a) VLA total intensity and (b) linear polarized intensity maps of the
 Boomerang PWN at 6\,GHz. 
 The beam size is $2\farcs3\times1\farcs8$.
 The white crosses mark the position of \psr.
 The radio images have a beam size of $2\farcs3\times1\farcs8$ FWHM and rms
 noise of 5\mjb. (c) same as (a) with different PWN features labeled. The arrow
 pointing west shows the pulsar spin axis direction inferred from the X-ray PWN
 \citep{ng04}. The arrow pointing southeast indicates the direction to the relic
 PWN \citep{kothes+01}. 
 (d) Exposure corrected \cxo\ X-ray image of the same
 field, showing the central pulsar J2229+6119 and the surrounding torus.}
    \label{fig:boomerang}
\end{figure*}

Figure~\ref{fig:boomerang}a shows the 6\,GHz total intensity map of the
Boomerang formed using the full bandwidth of 4\,GHz. Our high resolution images
reveal, for the first time, the detailed structure of the PWN. We discover
several bright radio features embedded in diffuse emission. In the inner PWN,
\psr\ is not detected in our observations, but there is a compact radio nebula,
which we call the ``core'', near the pulsar position. It has an elliptical shape
of $40''\times20''$ lying along the northeast-southwest direction and has
mean brightness temperature of $\sim$0.15\,K at 6\,GHz. The pulsar is located at
the southeastern edge of the core. Further southeast of the pulsar, there
exists a subluminous zone of $25\arcsec\times12\arcsec$ without obvious radio
emission. At the northern tip of the core, there is a hint of extended emission
towards southeast. If confirmed, then the overall morphology of the core would
also resemble a mini boomerang instead of merely elliptical.

In the north, there is a semicircular arc-like feature that encloses the entire
core. It has a width of 20\arcsec--30\arcsec\ with a sharp boundary in the north.
Its morphology closely follows that of the core, and there is a gap of $\sim30''$
between them (see Fig.~\ref{fig:boomerang}c). The arc can be roughly divided into
two parts. The northwestern part, which we refer to as the ``lobe'', is about
$150''\times50''$ in size, and it contains the peak emission of the radio PWN
with a brightness temperature of up to $\sim$0.3\,K at 6\,GHz. The northeastern part
of the arc is significantly fainter and smaller ($80''\times20''$) and is called
the ``tongue''\footnote{The ``tongue'' presented in \cite{kothes06} appears to have a different orientation to that identified in Fig.~\ref{fig:boomerang}.
We note that this is because images in \cite{kothes06} are plotted in the Galactic coordinate system, which is rotated by $\sim27^\circ$ compared to the equatorial coordinate system.} in a previous study \citep{kothes06}. All these bright features are
embedded in faint diffuse emission that fills the gap between the arc and the
core. 
It further extends and fades toward the southwest, roughly along the core orientation.
The full extent is uncertain because of the limited largest angular scales accessible to our observations (see also Figure~\ref{fig:full_fov} in the Appendix).

In Figure~\ref{fig:boomerang}d we show an X-ray image taken with the
\emph{Chandra X-ray Observatory} for comparison \citep{halpern+01a}. The X-ray
PWN exhibits a clear torus structure with a $\sim0.5'$ diameter \citep{ng04,ng08},
which is significantly smaller than the radio core and has a different orientation.
This is surrounded by faint diffuse X-ray emission of approximately circular
shape with $\sim2'$ diameter. The emission fills in the gap between the radio core and the arc. Its surface brightness drops radially outward and becomes undetectable near
the inner edge of the radio arc.

\subsection{Polarization}
The PI map of the Boomerang is shown in Figure~\ref{fig:boomerang}b. The
polarized emission well follows the total intensity. The core, the lobe, and the
tongue have high polarized fractions (PF) of $\sim$57\%, $\sim$60\% and
$\sim$45\%, respectively, and the average PF of the overall PWN is $\sim$50\%.
These values are higher than the previously reported value at a similar
frequency \citep[$37\pm5$\% at 4.85\,GHz;][]{kothes06}, likely due to reduced beam depolarization in our high-resolution image.
\begin{figure}
 \centering
 \includegraphics[width=0.47\textwidth]{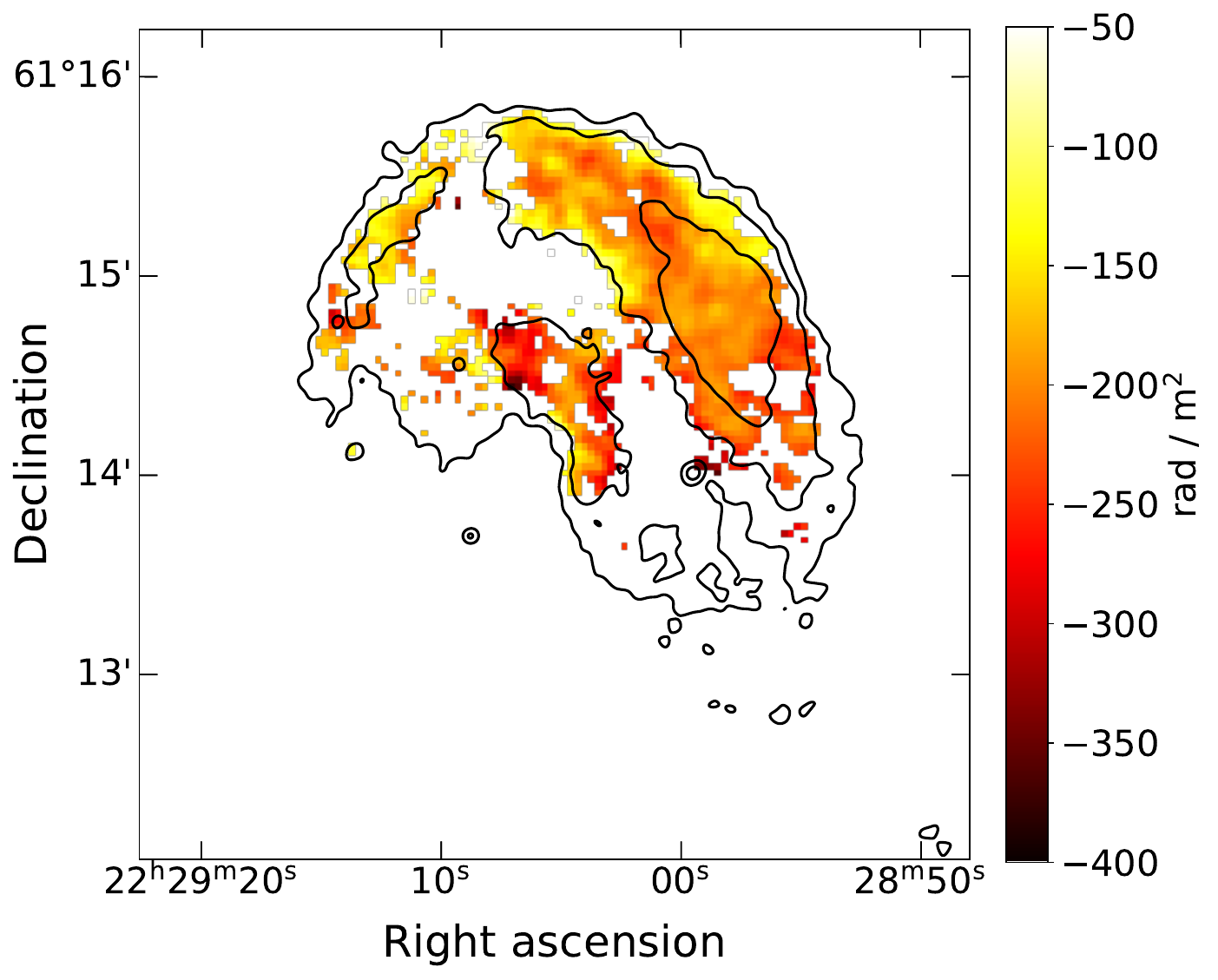}
 \caption{RM map of the Boomerang PWN overlaid with the 6\,GHz total intensity
 contours. 
 The angular resolution of the RM map is $5\farcs0\times5\farcs0$. The RM uncertainties are $<$45\,rad\,m$^{-2}$.}
 \label{fig:rm}
\end{figure}
\begin{figure}
 \centering
 \includegraphics[width=0.47\textwidth]{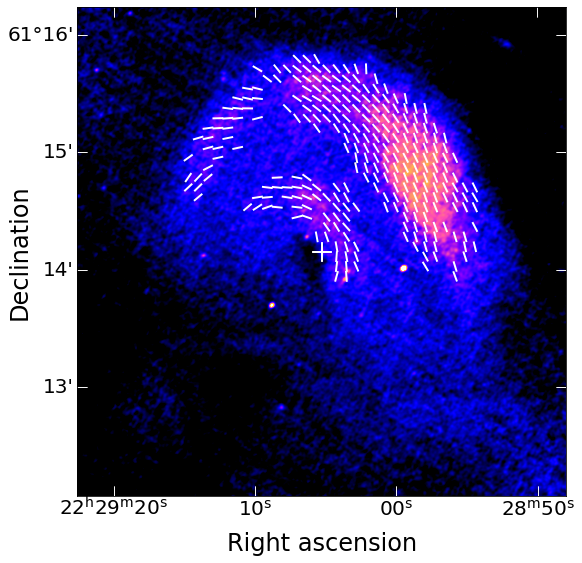}
 \caption{Total intensity map same as Figure~\ref{fig:boomerang}a (beam size $2\farcs3\times1\farcs8$), overlaid with
 unit vectors that show the intrinsic $B$-field orientation after corrected for
 Faraday rotation. 
 The $B$-field orientation is derived from the subband images of beam size $5\farcs0\times5\farcs0$. The vectors are only plotted for regions with Stokes I
 signal-to-noise ratio $>3$ and PA-fit uncertainty $<15\arcdeg$. The white cross
 marks the position of \psr.}
 \label{fig:mag}
\end{figure}

\begin{figure*}
\centering
 \includegraphics[width=0.85\textwidth]{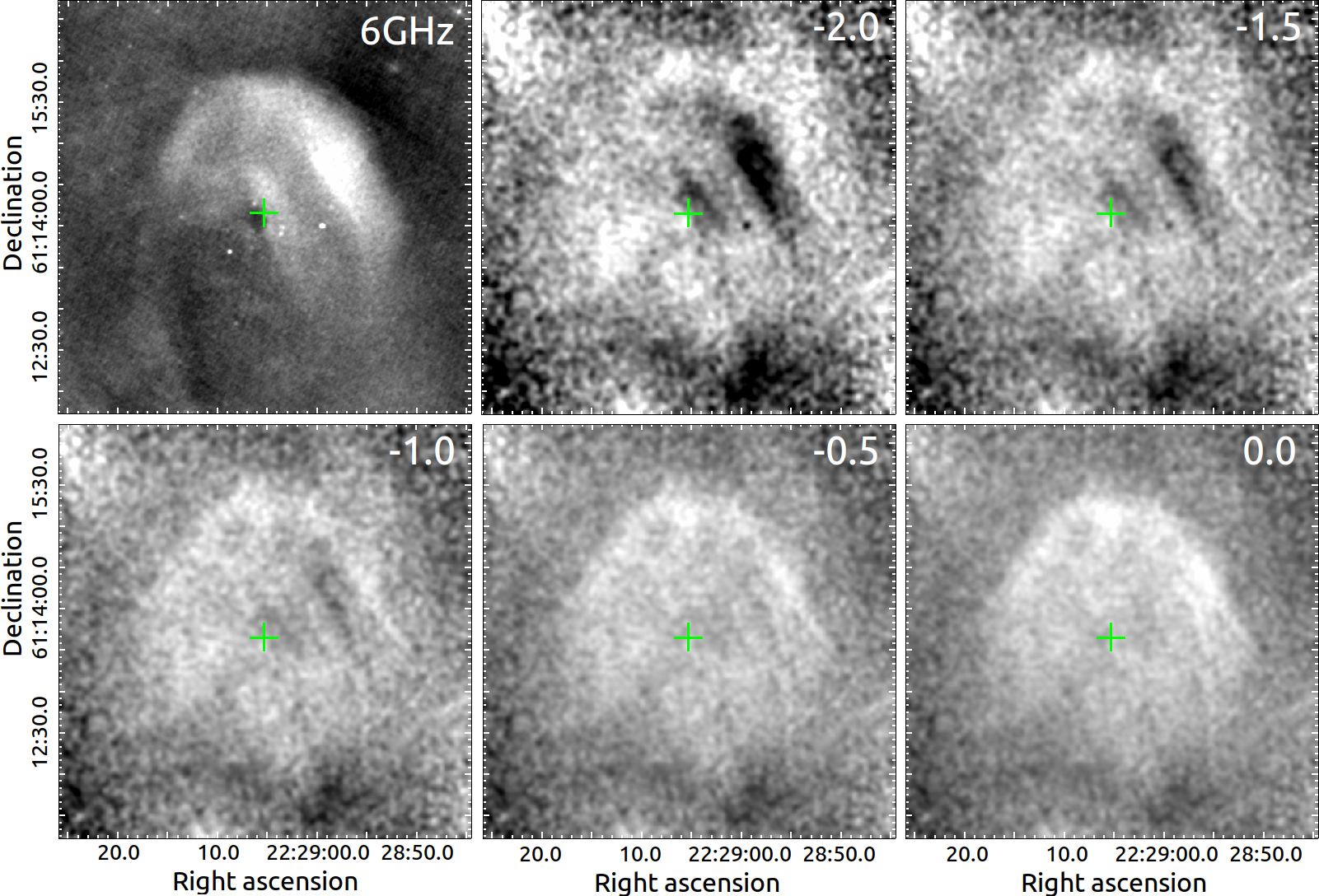}
 \caption{Selected spectral tomography images for the Boomerang between the
 4.5\,GHz and 7.5\,GHz subband total intensity images. The top left panel shows
 the full band intensity image as Fig.~\ref{fig:boomerang}a to illustrate the
 PWN features. For other panels, the trial spectral index $\alpha_t$ is labeled
 in top right. The cross marks the position of \psr.}
 \label{fig:tomo}
\end{figure*}

Figure~\ref{fig:rm} presents the RM map of the nebula. The RM value ranges from
$-380$ to $-20$\,rad\,m$^{-2}$ across the PWN, and the measurement uncertainties
are better than 45\,rad\,m$^{-2}$. Most region of the PWN has roughly constant RM
between $-250$ and $-150$\,rad\,m$^{-2}$. This is consistent with the values of
$\sim-200$\,rad\,m$^{-2}$ and $-187$\,rad\,m$^{-2}$ previously reported for the
nebula and the pulsar, respectively \citep{kothes06,ng+20}. The core, the lobe,
and the tongue have slightly different RM values ($-230$, $-180$, and
$-160\,$rad\,m$^{-2}$, respectively), but the discrepancy is not significant
given the measurement uncertainty. 

Figure~\ref{fig:mag} shows the intrinsic magnetic field orientation
(PA+$90\arcdeg$) of the Boomerang. We find a highly ordered field structure,
compatible with the large PF observed. The $B$-field in the core has a toroidal
configuration, consistent with previous findings \citep{kothes06} and the
prediction from the canonical PWN model \citep{kc84,conto+99}. On the other
hand, the $B$-field of the lobe and the tongue are mostly unidirectional. The
former well follows the lobe's elongation, while the latter mainly lies along
the east-west direction, misaligned with the elongation of the tongue. The
overall magnetic field structure of the lobe and the tongue shows a general
toroidal geometry similar to the core but with slight distortion. For example,
we would expect the field lines to curve around the northern tip, which is however
not observed.

\subsection{Spectrum}
Since the largest angular scale recoverable by our observation is $\sim2'$ (see Sec.~\ref{sec:data}), which is smaller than the angular extent of the Boomerang PWN ($\sim3'$), our measurements can only provide lower limits on its flux density. These lower limits were obtained by subtracting an estimate of the background level from the measured flux. We measured the flux densities of the entire Boomerang and its individual components from the 6\,GHz total intensity map using the regions shown in Figure~\ref{fig:full_fov}. 
To estimate the lower limit conservatively, we scanned regions outside the PWN with a circular aperture approximately the size of the Boomerang and adopted the maximum enclosed flux as the background level. The corresponding background region is also shown in Figure~\ref{fig:full_fov}. We derive lower limits on the flux densities of approximately 50, 5, 25, and 5\,mJy for the entire PWN, the tongue, the lobe, and the core, respectively.
We estimate the
spectral index using spectral tomography \citep[see][and references therein]{gaensler+wallace03}.
We scaled the
subband intensity maps at 4.5 and 7.5\,GHz with trial spectral indices
$\alpha_t$ to form difference images. When a feature blends into the background,
$\alpha_t$ gives its spectral index.
Figure~\ref{fig:tomo} shows the tomography images
with a few representative $\alpha_t$. The core and the lobe show a flat spectrum
with an index $\alpha\approx -0.5$, but the tongue and the northwestern edge of the
lobe have steeper spectrum of $\alpha\approx-1.0$ and $-1.5$, respectively. Note
that these estimates could be subject to large uncertainties of $\Delta\alpha\sim
0.2$.

\section{Discussion} \label{sec:discuss}

\subsection{PWN structure} \label{sec:morph}
Our VLA observations reveal new radio features of the Boomerang PWN, including the core, lobe, and tongue. The core exhibits a radially decreasing surface brightness, consistent with declining magnetic field strength or particle density as the post-shock wind diffuses outward, similar to the X-ray morphology of the Boomerang \citep{liang+22}. The outer arc, comprising the lobe and tongue, was previously detected only at 1.42\,GHz \citep{kothes06}, but our improved sensitivity clearly reveals emission up to at least 6\,GHz.

While the lobe and tongue are well-established structures, it remains unclear whether the relatively faint core is a genuine physical feature or simply a surface-brightness fluctuation. Regardless, PWNe are generally brightest near the pulsar, where freshly accelerated particles are injected. The main puzzle in the Boomerang is that the radio peak is offset from the pulsar by at least 0.3\,pc (assuming $d=0.8$\,kpc). One possibility is that the lobe arises from the interaction between the pulsar wind and a nearby dense cloud. Previous radio observations identified an H{\sc i} shell north of the Boomerang that is directly interacting with the pulsar wind \citep{kothes+01}. This scenario does not depend on the nature of the core.

Alternatively, if the core marks the termination shock, the Boomerang resembles the double torus/arc morphology seen in the X-ray PWNe of the Crab and MSH 15$-$5\emph{2} \citep{weisskopf+00,gaensler+02}. However, such morphology is rarely observed in radio PWNe. In young PWNe, radio structures can be disrupted by Rayleigh--Taylor instabilities driven by the interaction between the pulsar wind and the surrounding medium \citep{Chevalier+11}. These instabilities may be suppressed in the Boomerang by the high-pressure environment created by the supernova reverse shock, as proposed for the Vela PWN \citep{kothes06}. Consistent with this picture, our high-resolution radio image shows no filamentary structures, further supporting the absence of such instabilities.

Finally, the non-detection of \psr\ in our 6\,GHz data is not unexpected. At
lower frequencies it has phase-averaged flux densities of $1.5\pm0.1$\,mJy and
$0.25\pm0.08$\,mJy at 350\,MHz and 1412\,MHz, respectively \citep{halpern+01a,
stovall+14}. These give a spectral index of $1.3\pm0.3$. An extrapolation to
6\,GHz with a simple power-law suggests $20\pm10\,\mu$Jy, which is comparable
with the noise level of our image. The actual flux density could be lower if
there is any spectral turnover between 1--6\,GHz, as those gigahertz-peaked
pulsars \citep{Kijak+21}.

\subsection{Magnetic Field} \label{sec:bfield}
There is a previous claim that the Boomerang has radially increasing RM
distribution, and a radial $B$-field geometry was proposed \citep{kothes06}. Our
high resolution RM map found no evidence for such RM variation
(Fig.~\ref{fig:rm}). We suspect that the large beam size in the previous work
($1\farcm15$) could lead to artificial features because of the complex interplay
between the field geometry, the source brightness, and the external Faraday
dispersion. 
Our result prefers a toroidal field
instead. It well explains the polarization structure seen the core (see
Fig.~\ref{fig:mag}). For the lobe and the tongue, the $B$-field is mostly
toroidal, albeit with slight distortion that could possibly be due to
environmental effects.

\begin{figure}
    \centering
    \includegraphics[width=0.42\textwidth]{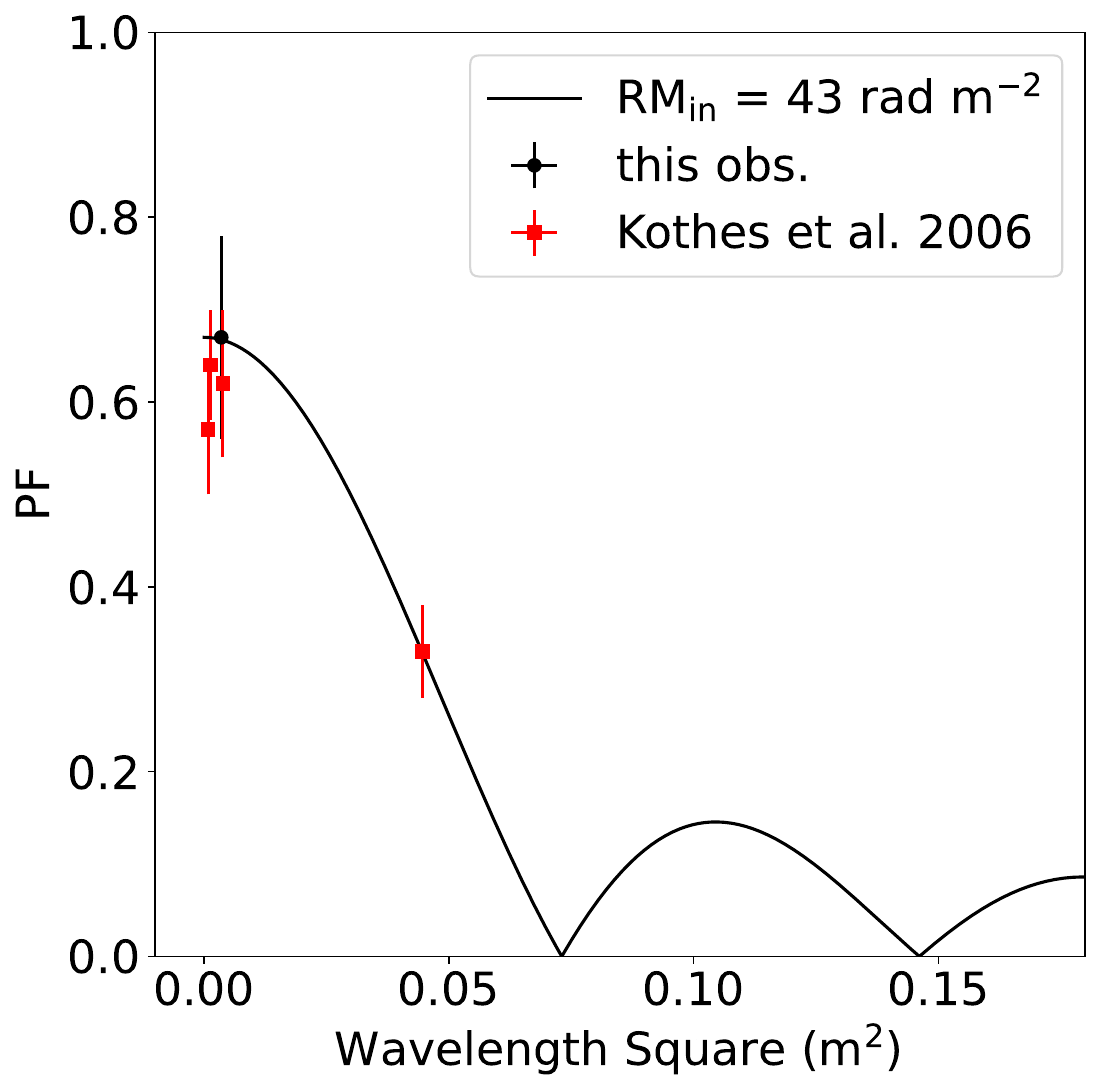}
    \caption{Observed PF versus the square of wavelength. The \citet{kothes06}
    data points are corrected for beam depolarization and the line shows the
    best-fit Burn slab model with $|$RM$_{\rm in}|$ $= 43$\,rad\,m$^{-2}$.  }
    \label{fig:pf_model}
\end{figure}

Our new study finds very high PF for the Boomerang. In particular, the lobe has
intrinsic PF of 67\% after accounting for the $\sim7$\% bandwidth
depolarization. This is very close to the theoretical maximum of $\sim$70\% for
synchrotron radiation\footnote{For synchrotron radiation, the maximum PF depends
on the spectral index $\alpha$ as PF$=(1+\alpha)/(5/3+\alpha)$. Adopting
$\alpha=0.59$ \citep{kothes06} gives PF$\approx$70\%.}, indicating a highly ordered
$B$-field with negligible depolarization. To compare with PF measured at other
frequencies in previous study, we smoothed our Stokes I, Q, and U maps to lower
resolution to determine the beam depolarization effect, assuming that the
relative brightness between different regions remains the same across frequency
bands. The PF values after correction are listed in
Table~\ref{tab:pf_corrected}. They are all near the synchrotron limit at 4.85,
8.35, and 10.45\,GHz, except at 1.42\,GHz which is only 33\%. These are plotted
verses the square of wavelength in Figure~\ref{fig:pf_model}.

\begin{deluxetable} {cccc} \label{tab:pf_corrected}
\caption{PF of the Boomerang PWN at different frequencies, from our new data and
previously reported values \citep{kothes06}. The last column lists the PF values
after correction for the beam depolarization effect.} \tablehead{
\colhead{Freq.\ (GHz)} & \colhead{resolution} & \colhead{measured
PF} & \colhead{corrected PF} }
\startdata
\multicolumn{4}{l}{This work} \\
6 & $\sim2''$ & $67\pm11$\% & $67\pm11$\% \\
\multicolumn{4}{l}{Previous observations} \\
1.42 & $50''$ & $31\pm5$\% & $33\pm5$\% \\
4.85 & $147''$ & $37\pm5$\% & $62\pm8$\% \\
8.35 & $84''$ & $52\pm5$\% & $64\pm6$\% \\
10.45 & $69''$ & $50\pm6$\% & $57\pm7$\% \\
\enddata
\end{deluxetable}

The low PF at 1.42\,GHz cannot be attributed to bandwidth depolarization, since
the observations have small bandwidth of 10--50\,MHz
\citep{Reich1997A&AS,Condon1998AJ,kothes+01}, which can induce at most 6\%
depolarization for RM=$-200$\,rad\,m$^{-2}$. We therefore believe that this is
due to Faraday effect, either non-uniform RM in the foreground (i.e.\ external
Faraday dispersion) or Faraday rotation within the PWN. The former effect can be
estimated by 
\begin{equation} \label{eq:EFD}
    {\rm PF} = {\rm PF}_{\rm max} e^{-2\sigma_{\rm RM}^2 \, \lambda^4},
\end{equation}
where $\sigma_{\rm RM}$ is the standard deviation of the RM values within an
observing beam \citep{burn66, Farnsworth2011AJ}. Comparing the observed PF of
33\% at 1.42\,GHz with PF$_{\rm max}=67\%$ implies $\sigma_{\rm RM} \approx
13\,$rad\,m$^{-2}$. 
We derived in Appendix~\ref{app:EFD}
that if the turbulence in the interstellar magnetic field follow the
Kolmogorov model, 
then $\sigma_{\rm RM}$ is related to the 
    characteristic $B$-field strength as
    $\sigma_{\rm RM} \, [{\rm rad}\,{\rm m}^{-2}] \approx \sqrt{0.51 B_{\rm ran} \, [\mu{\rm G}]+0.04}$
and
$\sigma_{\rm RM} \, [{\rm rad}\,{\rm m}^{-2}] \approx \sqrt{4.8B^2_{\rm ran} \, [\mu{\rm G}]+9.3 }$
for a distance of 0.8\,kpc and 8\,kpc, respectively.
This requires $B_{\rm ran} \approx 26\,\mu$G 
    or $B_{\rm ran}\approx5.8\,\mu$G
in the interstellar medium
(ISM) along the line of sight. 
The former is much higher than the typical strength in the ISM and the latter is mildly higher than the
average strength of $\sim$3--5\,$\mu$G.
We therefore conclude that this scenario is unlikely.
Note that our high resolution RM map has too large uncertainty
(45\,rad\,m$^{-2}$) for $\sigma_{\rm RM}$ estimate. Better measurements are
needed to determine the true value of $\sigma_{\rm RM}$ from observation.

We next discuss the case that the PWN medium is Faraday thick, such that
depolarization is caused by Faraday rotation within the emission volume.
Here we focus on the lobe only as it dominants the observed flux. We model it as
a Burn slab in the plane of the sky with uniform emissivity, density, and
magnetic field \citep{burn66}. The observed PF is given by
\begin{equation} \label{eq:burn}
 {\rm PF}={\rm PF}_{\rm max}\left|\frac{\sin{(\lambda^2{\rm RM}_{\rm in})}}
 {\lambda^2{\rm RM}_{\rm in}} \right|,
\end{equation}
where $\lambda$ is the observation wavelength and ${\rm RM}_{\rm in}$ is the
equivalent RM if the slab acts as a foreground Faraday screen \citep{burn66}.
Taking ${\rm PF}_{\rm max} = 67\%$, $|{\rm RM}_{\rm in}| \approx
43\,$rad\,m$^{-2}$ is needed  to explain the PF of 33\% at 1.42\,GHz. This is
illustrated in Figure~\ref{fig:pf_model}.

The internal RM depends on the slab thickness $l$, the magnetic field strength
along the line of sight $B\cos\theta$, and the cold electrons density $n_{e,{\rm
cold}}$ as
\begin{equation}
  {\rm RM}_{\rm in}\approx0.812n_{e,{\rm cold}} B l \cos\theta.
\end{equation}
We ignore relativistic electrons since they contribute very little to the
Faraday rotation \citep{jones_odell77}. $B$ can be estimated with the standard
equipartition assumption
\begin{equation}
    B_{\rm eq} = [6\pi c_{12} (1+k) L \Phi^{-1} V^{-1}]^{2/7}  (\sin\theta)^{-3/7},
\end{equation}
where $c_{12}$ is a constant \citep{pac70,govoni+04}. We assume the
ion-to-electron energy ratio $k=0$ and the volume filling factor $\Phi = 1$.
We measured that the lower limit of the flux density of the lobe is 25\,mJy at 6\,GHz, while the upper limit can be conservatively set by the flux density of the entire PWN, which is around 65\,mJy \citep{kothes06}.
Assuming a spectral index of $-0.5$, we obtain
synchrotron luminosity 
$L\approx(3$--$8)\times10^{33}\,d^2_{0.8}$\,erg\,s$^{-1}$,
assuming the emission is in the range $6\times10^9$ to $10^{18}$\,Hz
\citep{liang+22} at the distance $d = 0.8\,d_{0.8}\,$kpc.
We take $l=0.2$\,pc, same as the width of the lobe. This gives the
emission volume $V \approx 6\times10^{53}\,d_{0.8}^3$\,cm$^3$,
and hence
\begin{equation} \label{eq:b_eq_num}
    B_{\rm eq} \sim(50\textrm{--}70)\,
    (\sin\theta)^{-3/7} 
    d_{0.8}^{-2/7}\,\mu{\rm G} 
\end{equation}
and 
\begin{equation} \label{eq:n_ecold_num}
    n_{\rm e, cold} \sim
    (3\textrm{--}6)\, 
    (\sin\theta)^{3/7}\,(\cos\theta)^{-1}\, 
    d_{0.8}^{-5/7}\,{\rm cm}^{-3},
\end{equation}
where the range arises from the uncertainty in the flux measurement.
Since the inclination angle is unknown,
    we estimate the order-of-magnitude values of the magnetic field strength and electron density by assuming a median inclination angle
    of $60^\circ$, which gives
    $B_{\rm eq} \sim (50\textrm{--}70)\,d_{0.8}^{-2/7}\,\mu$G and 
    $n_{\rm e,cold}\sim(5\textrm{--}10)\,d^{-5/7}_{0.8}\,$cm$^{-3}$.
The former
is compatible with estimates from previous study \citep[e.g.,][]{liang+22,
Pope2024ApJ}, and the latter is feasible if the pulsar wind is mixing with the
nearby H{\sc i} cloud.
    
\section{Conclusion} \label{sec:conclusion}

We present a high resolution radio study of the Boomerang using VLA observations
at 6\,GHz. We discover that the radio PWN consists of a compact core
surrounding the pulsar and a bright lobe with a tongue. The latter two form an
outer arc of $\sim2'$ diameter. Such structure is rather uncommon among PWNe. We
suggest that the arc could be either intrinsic due to the flow structure, or due
to the environment, either interaction with the supernova reverse shock or with
the surrounding dense cloud. Our polarization measurement reveals a nebular
$B$-field that is mostly toroidal, consistent with what conventional theories
predict. We also find a high PF at 6\,GHz close to the synchrotron limit,
implying a highly ordered $B$-field structure. On the other hand, the PF at
1.42\,GHz is significantly lower. We attribute the discrepancy to Faraday
rotation within the emission volume. 
We therefore estimate that the magnetic field strength is in the order of 50--70\,$\mu$G and 25--40\,$\mu$G,
for $d=0.8\,$kpc and $d=8\,$kpc, respectively, assuming a median inclination angle of $60^\circ$.

\begin{acknowledgments}

We thank Roland Kothes for providing us the data of  their previous observations
on the Boomerang PWN. PCWL is supported by a UCL Graduate Research Scholarship
and a UCL Overseas Research Scholarship. 
C.-Y. N. and S. Zhang are supported by GRF grants of the Hong Kong Government
under HKU 7301723 and HKU 17304524. The National Radio Astronomy Observatory is
a facility of the National Science Foundation operated under cooperative
agreement by Associated Universities, Inc.
\end{acknowledgments}

\facilities{VLA}

\software{CASA \citep{casa}}

\appendix 

\renewcommand{\theequation}{A\arabic{equation}}
\renewcommand{\theHequation}{appendix.A\arabic{equation}}

\section{External Faraday Dispersion} \label{app:EFD}

To derive the expected external Faraday dispersion,
    we first find the suitable RM structure function that describes the Boomerang PWN.
The RM structure function is defined as 
\begin{equation}
    D_{\rm RM}(\delta \theta) =
    \langle [{\rm RM}(\theta) - {\rm RM}(\theta+\delta\theta)]^2 \rangle_\theta \ .
\end{equation}
It describes the mean squared difference of RM
    at two positions given a angular separation of $\delta \theta$.
    
The RM structure function is derived by 
    \cite{minter+96} assuming a turbulent 
    interstellar medium
    that follows the Kolmogorov model,
    and we find the values of the parameters 
    following the procedures in \cite{lai+22}.
The RM structure function can be expressed as
\begin{equation} \label{eq:rm structure}
\begin{split}
    D_{\rm RM} (\delta \theta)= &\Bigg\{ 251.226 \Bigg[ \left( \frac{n_{e}}{0.1\,{\rm cm}^{-3}} \right)^2 \left( \frac{C_B^2}{10^{-13}\,{\rm m}^{-2/3}\,\mu{\rm G}^2} \right) \\ 
    &+ \left( \frac{B_{\|}}{\mu{\rm G}} \right)^2 \left( \frac{C_n^2}{10^{-3}\,{\rm m}^{-20/3}} \right) \Bigg] \\
    &+ 23.043 \left( \frac{C_n^2}{10^{-3}\,{\rm m}^{-20/3}} \right) \\
    &\times \left( \frac{C_B^2}{10^{-13}\,{\rm m}^{-2/3}\,\mu{\rm G}^2} \right) \left( \frac{l_0}{\rm pc} \right)^{2/3} \Bigg\} \\
    &\times \left( \frac{d}{\rm kpc} \right)^{8/3} \left( \frac{\delta\theta}{\rm deg} \right)^{5/3},
\end{split}
\end{equation}
where $n_e$ is the mean density of the medium, $B_\parallel$ is the mean magnetic field strength of the line-of-sight component, $d$ is the distance to the source, $l_0$ is the outer scale of the Kolmogorov turbulence, $C_n^2$ and $C_B^2$ is related to the  level of fluctuation in density and magnetic field, respectively \citep{minter+96}. In particular, $C_B^2$ can be expressed as $C_B^2 = 5.2B_{\rm ran}^2 (l_0)^{-2/3}  \times 10^{-13}\,$\,m$^{-2/3}\,\mu$G$^2$, where $B_{\rm ran}$ is the strength of the random $B$-field.
    
From observations of the host pulsar of  the Boomerang PWN PSR J2229+6114, we know
    ${\rm DM} = n_e d \approx 205\,$pc\,cm$^{-3}$ \citep{McEwen2020ApJ} and
    $|{\rm RM}| \approx 0.812 \, n_e |B_\parallel| d = 187\,$rad\,m$^{-2}$ \citep{ng+20}.
For a distance of $d = 0.8\,$kpc \citep{kothes+01},
    $n_e \approx 0.26\,$cm$^{-3}$ and 
    $|B_\parallel| \approx 1.1\,\mu$G. 
We further adopt $l_0 = 3.6\,$pc and $C_n^2 = 10^{-3}\,$m$^{-20/3}$ as suggested in
    \cite{minter+96}.
$B_{\rm ran}$ of the interstellar medium
    varies quite a lot depending on the
    location, but is in general smaller than
    10\,$\mu$G 
    \citep{minter+96, haverkorn+08}.
Putting the values back into Eq.~\ref{eq:rm structure},
    we get 
\begin{equation} \label{eq:rm structure2}
    D_{\rm RM}(\delta \theta) \approx \Bigg[ 2.33 \left( \frac{B_{\rm ran}}{\mu{\rm G}} \right)^2 + 0.18 \Bigg] 
    \left( \frac{\delta\theta}{1'} \right)^{5/3} \,{\rm rad}^2\,{\rm m}^{-4}.
\end{equation}

Given a circle with angular size $r$, the expected value of $D_{\rm RM}$ between two randomly selected points within the circle can be expressed as
\begin{eqnarray}
    \langle D_{\rm RM} \rangle_r &= &
    \langle ({\rm RM}_i - {\rm RM}_j)^2 \rangle_r  \nonumber \\
    &=& 2 \langle {\rm RM}^2 \rangle_r - 2 \langle {\rm RM} \rangle^2_r  \nonumber\\
    &=& 2 \sigma_{\rm RM}^2,
\end{eqnarray}
where $\sigma_{\rm RM}$ is the standard deviation of RM within a beam that we
want to find for Eq.~\ref{eq:EFD}. With Eq.~\ref{eq:rm structure2}, we calculate
$\langle D_{\rm RM} \rangle_r$ using Monte Carlo simulation and found 
    $\langle D_{\rm RM} \rangle_{r = 50''} \approx 0.51 B_{\rm ran}^2$.
Therefore, we obtain 
\begin{equation}
    \sigma_{\rm RM} \, [{\rm rad}\,{\rm m}^{-2}] \approx 
    \sqrt{0.25 B^2_{\rm ran} \, [\mu{\rm G}]+0.02} ,
\end{equation}
for $d = 0.8\,$kpc.
Repeating the above calculation assuming
    a larger distance $L = 8\,$kpc,
    we obtain
\begin{equation}
    \sigma_{\rm RM} \, [{\rm rad}\,{\rm m}^{-2}] \approx \sqrt{4.8B^2_{\rm ran} \, [\mu{\rm G}]+9.3 } \ ,
\end{equation}
with $n_e \approx 0.026\,$cm$^{-3}$
    and 
    $|B_\parallel| \approx 1.1\,\mu$G.

\section{Supplementary images}
\begin{figure*}
    \centering
    \includegraphics[width=0.7\linewidth]{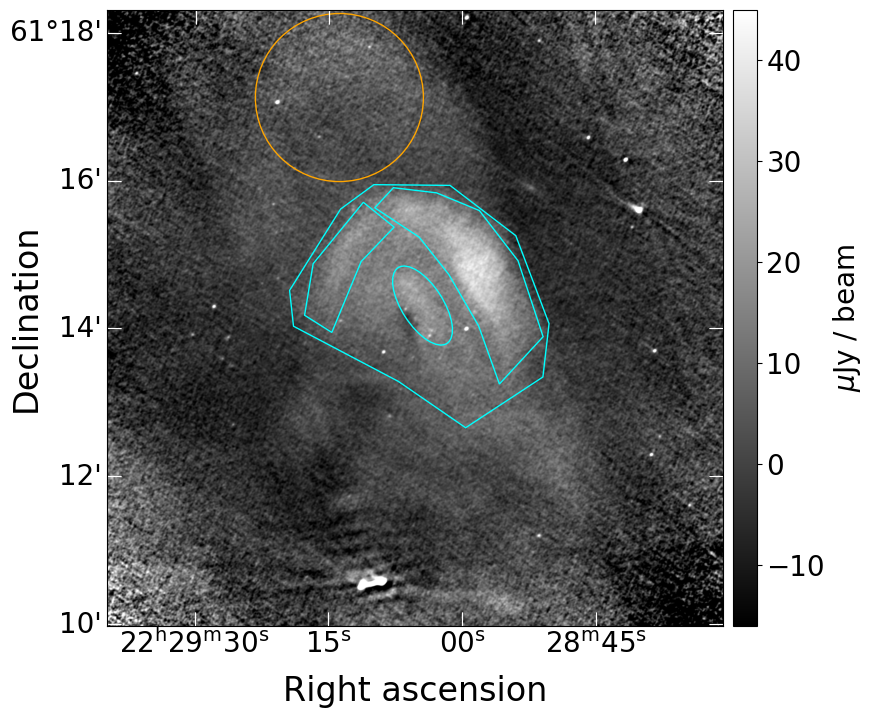}
    \caption{
    Total intensity map of the Boomerang PWN at 6\,GHz.
    It is the same image as Figure~\ref{fig:boomerang}a with a larger field of view and a different color stretch to show the background level. The background is highly non-uniform likely due to poor $u$-$v$ coverage of our observations.
    The outlined regions are used for flux measurements. 
    From the largest to the smallest regions, they are the entire Boomerang PWN, the lobe, the tongue, and the core.
    The circular region is used for background subtraction in flux measurements.}
    \label{fig:full_fov}
\end{figure*}

\bibliography{pwn} 

@ARTICLE{Pineault2000AJ,
       author = {{Pineault}, Serge and {Joncas}, Gilles},
        title = "{G106.3+2.7: A Supernova Remnant in a Late Stage of Evolution}",
      journal = {\aj},
         year = 2000,
        month = dec,
       volume = {120},
       number = {6},
        pages = {3218-3225},
          doi = {10.1086/316863},
       adsurl = {https://ui.adsabs.harvard.edu/abs/2000AJ....120.3218P}
}

@incollection{Amato2024PulsarAcceleration,
  author    = {Elena Amato},
  title     = {Particle Acceleration in Pulsars and Pulsar Wind Nebulae},
  booktitle = {Foundations of Cosmic Ray Astrophysics},
  editor    = {Felix A. Aharonian and Elena Amato and Pasquale Blasi and Carmelo Evoli},
  publisher = {IOS Press},
  year      = {2024},
  pages     = {183--221},
  doi       = {10.3254/ENFI240006}
}

@ARTICLE{Kothes2006A&A,
       author = {{Kothes}, R. and {Fedotov}, K. and {Foster}, T.~J. and {Uyan{\i}ker}, B.},
        title = "{A catalogue of Galactic supernova remnants from the Canadian Galactic plane survey. I. Flux densities, spectra, and polarization characteristics}",
      journal = {\aap},
         year = 2006,
        month = oct,
       volume = {457},
       number = {3},
        pages = {1081-1093},
          doi = {10.1051/0004-6361:20065062},
       adsurl = {https://ui.adsabs.harvard.edu/abs/2006A&A...457.1081K}
}

@ARTICLE{Liu23ApJ,
       author = {{Liu}, Y.~H. and {Ng}, C. -Y. and {Dodson}, R.},
        title = "{Radio Study of the Pulsar Wind Nebula Powered by PSR B1706-44}",
      journal = {\apj},
         year = 2023,
        month = mar,
       volume = {945},
       number = {1},
          eid = {82},
        pages = {82},
          doi = {10.3847/1538-4357/acb20d},
archivePrefix = {arXiv},
       eprint = {2301.13651},
 primaryClass = {astro-ph.HE},
       adsurl = {https://ui.adsabs.harvard.edu/abs/2023ApJ...945...82L}
}

@ARTICLE{Pope2024ApJ,
       author = {{Pope}, I. and {Mori}, K. and {Abdelmaguid}, M. and {Gelfand}, J.~D. and {Reynolds}, S.~P. and {Safi-Harb}, S. and {Hailey}, C.~J. and {An}, H. and {Bangale}, P. and {Batista}, P. and {Benbow}, W. and {Buckley}, J.~H. and {Capasso}, M. and {Christiansen}, J.~L. and {Chromey}, A.~J. and {Falcone}, A. and {Feng}, Q. and {Finley}, J.~P. and {Foote}, G.~M. and {Gallagher}, G. and {Hanlon}, W.~F. and {Hanna}, D. and {Hervet}, O. and {Holder}, J. and {Humensky}, T.~B. and {Jin}, W. and {Kaaret}, P. and {Kertzman}, M. and {Kieda}, D. and {Kleiner}, T.~K. and {Korzoun}, N. and {Krennrich}, F. and {Kumar}, S. and {Lang}, M.~J. and {Maier}, G. and {McGrath}, C.~E. and {Mooney}, C.~L. and {Moriarty}, P. and {Mukherjee}, R. and {O'Brien}, S. and {Ong}, R.~A. and {Park}, N. and {Patel}, S.~R. and {Pfrang}, K. and {Pohl}, M. and {Pueschel}, E. and {Quinn}, J. and {Ragan}, K. and {Reynolds}, P.~T. and {Roache}, E. and {Sadeh}, I. and {Saha}, L. and {Sembroski}, G.~H. and {Tak}, D. and {Tucci}, J.~V. and {Weinstein}, A. and {Williams}, D.~A. and {Woo}, J. and {VERITAS Collaboration}},
        title = "{A Multiwavelength Investigation of PSR J2229+6114 and its Pulsar Wind Nebula in the Radio, X-Ray, and Gamma-Ray Bands}",
      journal = {\apj},
         year = 2024,
        month = jan,
       volume = {960},
       number = {1},
          eid = {75},
        pages = {75},
          doi = {10.3847/1538-4357/ad0120},
archivePrefix = {arXiv},
       eprint = {2310.04512},
 primaryClass = {astro-ph.HE},
       adsurl = {https://ui.adsabs.harvard.edu/abs/2024ApJ...960...75P}
}

@ARTICLE{McEwen2020ApJ,
       author = {{McEwen}, A.~E. and {Spiewak}, R. and {Swiggum}, J.~K. and {Kaplan}, D.~L. and {Fiore}, W. and {Agazie}, G.~Y. and {Blumer}, H. and {Chawla}, P. and {DeCesar}, M. and {Kaspi}, V.~M. and {Kondratiev}, V.~I. and {LaRose}, M. and {Levin}, L. and {Lynch}, R.~S. and {McLaughlin}, M. and {Mingyar}, M. and {Noori}, H. Al and {Ransom}, S.~M. and {Roberts}, M.~S.~E. and {Schmiedekamp}, A. and {Schmiedekamp}, C. and {Siemens}, X. and {Stairs}, I. and {Stovall}, K. and {Surnis}, M. and {van Leeuwen}, J.},
        title = "{The Green Bank North Celestial Cap Pulsar Survey. V. Pulsar Census and Survey Sensitivity}",
      journal = {\apj},
         year = 2020,
        month = apr,
       volume = {892},
       number = {2},
          eid = {76},
        pages = {76},
          doi = {10.3847/1538-4357/ab75e2},
archivePrefix = {arXiv},
       eprint = {1909.11109},
 primaryClass = {astro-ph.HE},
       adsurl = {https://ui.adsabs.harvard.edu/abs/2020ApJ...892...76M}
}

@ARTICLE{Reich1997A&AS,
       author = {{Reich}, P. and {Reich}, W. and {Furst}, E.},
        title = "{The Effelsberg 21 CM radio continuum survey of the Galactic plane between L = 95.5 deg and L = 240 deg.}",
      journal = {\aaps},
         year = 1997,
        month = dec,
       volume = {126},
        pages = {413-435},
          doi = {10.1051/aas:1997274},
       adsurl = {https://ui.adsabs.harvard.edu/abs/1997A&AS..126..413R}
}

@ARTICLE{Condon1998AJ,
       author = {{Condon}, J.~J. and {Cotton}, W.~D. and {Greisen}, E.~W. and {Yin}, Q.~F. and {Perley}, R.~A. and {Taylor}, G.~B. and {Broderick}, J.~J.},
        title = "{The NRAO VLA Sky Survey}",
      journal = {\aj},
         year = 1998,
        month = may,
       volume = {115},
       number = {5},
        pages = {1693-1716},
          doi = {10.1086/300337},
       adsurl = {https://ui.adsabs.harvard.edu/abs/1998AJ....115.1693C}
}

@ARTICLE{Farnsworth2011AJ,
       author = {{Farnsworth}, Damon and {Rudnick}, Lawrence and {Brown}, Shea},
        title = "{Integrated Polarization of Sources at {\ensuremath{\lambda}} \raisebox{-0.5ex}\textasciitilde 1 m and New Rotation Measure Ambiguities}",
      journal = {\aj},
         year = 2011,
        month = jun,
       volume = {141},
       number = {6},
          eid = {191},
        pages = {191},
          doi = {10.1088/0004-6256/141/6/191},
archivePrefix = {arXiv},
       eprint = {1103.4149},
 primaryClass = {astro-ph.CO},
       adsurl = {https://ui.adsabs.harvard.edu/abs/2011AJ....141..191F}
}

@ARTICLE{Kijak+21,
       author = {{Kijak}, J. and {Basu}, R. and {Lewandowski}, W. and {Ro{\.z}ko}, K.},
        title = "{Low-frequency Flux Density Measurements and Pulsars with GHz-peaked Spectra}",
      journal = {\apj},
         year = 2021,
        month = dec,
       volume = {923},
       number = {2},
          eid = {211},
        pages = {211},
          doi = {10.3847/1538-4357/ac3082},
       adsurl = {https://ui.adsabs.harvard.edu/abs/2021ApJ...923..211K}
}

@ARTICLE{Olmi23,
       author = {{Olmi}, Barbara and {Bucciantini}, Niccol{\`o}},
        title = "{The Dawes Review 11: From young to old: The evolutionary path of Pulsar Wind Nebulae}",
      journal = {\pasa},
         year = 2023,
        month = mar,
       volume = {40},
          eid = {e007},
        pages = {e007},
          doi = {10.1017/pasa.2023.5},
archivePrefix = {arXiv},
       eprint = {2301.12903},
 primaryClass = {astro-ph.HE},
       adsurl = {https://ui.adsabs.harvard.edu/abs/2023PASA...40....7O},
}

@ARTICLE{liang+22,
       author = {{Liang}, Xuan-Han and {Li}, Chao-Ming and {Wu}, Qi-Zuo and {Pan}, Jia-Shu and {Liu}, Ruo-Yu},
        title = "{A PeVatron Candidate: Modeling the Boomerang Nebula in X-ray Band}",
      journal = {Universe},
         year = 2022,
        month = oct,
       volume = {8},
       number = {10},
        pages = {547},
          doi = {10.3390/universe8100547},
archivePrefix = {arXiv},
       eprint = {2209.03809},
 primaryClass = {astro-ph.HE},
       adsurl = {https://ui.adsabs.harvard.edu/abs/2022Univ....8..547L}
}

@ARTICLE{gaensler+02,
       author = {{Gaensler}, B.~M. and {Arons}, J. and {Kaspi}, V.~M. and {Pivovaroff}, M.~J. and {Kawai}, N. and {Tamura}, K.},
        title = "{Chandra Imaging of the X-Ray Nebula Powered by Pulsar B1509-58}",
      journal = {\apj},
         year = 2002,
        month = apr,
       volume = {569},
       number = {2},
        pages = {878-893},
          doi = {10.1086/339354},
archivePrefix = {arXiv},
       eprint = {astro-ph/0110454},
 primaryClass = {astro-ph},
       adsurl = {https://ui.adsabs.harvard.edu/abs/2002ApJ...569..878G}
}

@ARTICLE{stovall+14,
       author = {{Stovall}, K. and {Lynch}, R.~S. and {Ransom}, S.~M. and {Archibald}, A.~M. and {Banaszak}, S. and {Biwer}, C.~M. and {Boyles}, J. and {Dartez}, L.~P. and {Day}, D. and {Ford}, A.~J. and {Flanigan}, J. and {Garcia}, A. and {Hessels}, J.~W.~T. and {Hinojosa}, J. and {Jenet}, F.~A. and {Kaplan}, D.~L. and {Karako-Argaman}, C. and {Kaspi}, V.~M. and {Kondratiev}, V.~I. and {Leake}, S. and {Lorimer}, D.~R. and {Lunsford}, G. and {Martinez}, J.~G. and {Mata}, A. and {McLaughlin}, M.~A. and {Roberts}, M.~S.~E. and {Rohr}, M.~D. and {Siemens}, X. and {Stairs}, I.~H. and {van Leeuwen}, J. and {Walker}, A.~N. and {Wells}, B.~L.},
        title = "{The Green Bank Northern Celestial Cap Pulsar Survey. I. Survey Description, Data Analysis, and Initial Results}",
      journal = {\apj},
         year = 2014,
        month = aug,
       volume = {791},
       number = {1},
          eid = {67},
        pages = {67},
          doi = {10.1088/0004-637X/791/1/67},
archivePrefix = {arXiv},
       eprint = {1406.5214},
 primaryClass = {astro-ph.HE},
       adsurl = {https://ui.adsabs.harvard.edu/abs/2014ApJ...791...67S}
}

@ARTICLE{lai+22,
       author = {{Lai}, Paul C.~W. and {Ng}, C. -Y. and {Bucciantini}, Niccolo'},
        title = "{Magnetic Field Structure and Faraday Rotation of the Plerionic Supernova Remnant G21.5-0.9}",
      journal = {\apj},
         year = 2022,
        month = may,
       volume = {930},
       number = {1},
          eid = {1},
        pages = {1},
          doi = {10.3847/1538-4357/ac63b1},
archivePrefix = {arXiv},
       eprint = {2109.04156},
 primaryClass = {astro-ph.HE},
       adsurl = {https://ui.adsabs.harvard.edu/abs/2022ApJ...930....1L}
}

@ARTICLE{ge+21,
       author = {{Ge}, Chong and {Liu}, Ruo-Yu and {Niu}, Shu and {Chen}, Yang and {Wang}, Xiang-Yu},
        title = "{Revealing a peculiar supernova remnant G106.3+2.7 as a petaelectronvolt proton accelerator with X-ray observations}",
      journal = {The Innovation},
         year = 2021,
        month = may,
       volume = {2},
          eid = {100118},
        pages = {100118},
          doi = {10.1016/j.xinn.2021.100118},
archivePrefix = {arXiv},
       eprint = {2012.11531},
 primaryClass = {astro-ph.HE},
       adsurl = {https://ui.adsabs.harvard.edu/abs/2021Innov...200118G}
}

@ARTICLE{ng+20,
       author = {{Ng}, C. and {Pandhi}, A. and {Naidu}, A. and {Fonseca}, E. and {Kaspi}, V.~M. and {Masui}, K.~W. and {Mckinven}, R. and {Renard}, A. and {Scholz}, P. and {Stairs}, I.~H. and {Tendulkar}, S.~P. and {Vanderlinde}, K.},
        title = "{Faraday rotation measures of Northern hemisphere pulsars using CHIME/Pulsar}",
      journal = {\mnras},
         year = 2020,
        month = aug,
       volume = {496},
       number = {3},
        pages = {2836-2848},
          doi = {10.1093/mnras/staa1658},
archivePrefix = {arXiv},
       eprint = {2006.06538},
 primaryClass = {astro-ph.IM},
       adsurl = {https://ui.adsabs.harvard.edu/abs/2020MNRAS.496.2836N}
}

@ARTICLE{yang+22,
       author = {{Yang}, Chuyuan and {Zeng}, Houdun and {Bao}, Biwen and {Zhang}, Li},
        title = "{Possible hadronic origin of TeV photon emission from SNR G106.3+2.7}",
      journal = {\aap},
         year = 2022,
        month = feb,
       volume = {658},
          eid = {A60},
        pages = {A60},
          doi = {10.1051/0004-6361/202141850},
archivePrefix = {arXiv},
       eprint = {2106.02629},
 primaryClass = {astro-ph.HE},
       adsurl = {https://ui.adsabs.harvard.edu/abs/2022A&A...658A..60Y}
}

@ARTICLE{liu+20,
       author = {{Liu}, Siming and {Zeng}, Houdun and {Xin}, Yuliang and {Zhu}, Hui},
        title = "{Hadronic versus Leptonic Models for {\ensuremath{\gamma}}-Ray Emission from VER J2227+608}",
      journal = {\apjl},
         year = 2020,
        month = jul,
       volume = {897},
       number = {2},
          eid = {L34},
        pages = {L34},
          doi = {10.3847/2041-8213/ab9ff2},
archivePrefix = {arXiv},
       eprint = {2006.14946},
 primaryClass = {astro-ph.HE},
       adsurl = {https://ui.adsabs.harvard.edu/abs/2020ApJ...897L..34L}
}

@ARTICLE{halpern+01a,
       author = {{Halpern}, J.~P. and {Camilo}, F. and {Gotthelf}, E.~V. and {Helfand}, D.~J. and {Kramer}, M. and {Lyne}, A.~G. and {Leighly}, K.~M. and {Eracleous}, M.},
        title = "{PSR J2229+6114: Discovery of an Energetic Young Pulsar in the Error Box of the EGRET Source 3EG J2227+6122}",
      journal = {\apjl},
         year = 2001,
        month = may,
       volume = {552},
       number = {2},
        pages = {L125-L128},
          doi = {10.1086/320347},
archivePrefix = {arXiv},
       eprint = {astro-ph/0104109},
 primaryClass = {astro-ph},
       adsurl = {https://ui.adsabs.harvard.edu/abs/2001ApJ...552L.125H}
}

@ARTICLE{boom_lhaaso_2021,
       author = {{Cao}, Zhen and {Aharonian}, F.~A. and {An}, Q. and {Axikegu}, Bai, L.~X. and {Bai}, Y.~X. and {Bao}, Y.~W. and {Bastieri}, D. and {Bi}, X.~J. and {Bi}, Y.~J. and {Cai}, H. and {Cai}, J.~T. and {Cao}, Zhe and {Chang}, J. and {Chang}, J.~F. and {Chang}, X.~C. and {Chen}, B.~M. and {Chen}, J. and {Chen}, L. and {Chen}, Liang and {Chen}, Long and {Chen}, M.~J. and {Chen}, M.~L. and {Chen}, Q.~H. and {Chen}, S.~H. and {Chen}, S.~Z. and {Chen}, T.~L. and {Chen}, X.~L. and {Chen}, Y. and {Cheng}, N. and {Cheng}, Y.~D. and {Cui}, S.~W. and {Cui}, X.~H. and {Cui}, Y.~D. and {Dai}, B.~Z. and {Dai}, H.~L. and {Dai}, Z.~G. and {Danzengluobu} and {della Volpe}, D. and {D'Ettorre Piazzoli}, B. and {Dong}, X.~J. and {Fan}, J.~H. and {Fan}, Y.~Z. and {Fan}, Z.~X. and {Fang}, J. and {Fang}, K. and {Feng}, C.~F. and {Feng}, L. and {Feng}, S.~H. and {Feng}, Y.~L. and {Gao}, B. and {Gao}, C.~D. and {Gao}, Q. and {Gao}, W. and {Ge}, M.~M. and {Geng}, L.~S. and {Gong}, G.~H. and {Gou}, Q.~B. and {Gu}, M.~H. and {Guo}, J.~G. and {Guo}, X.~L. and {Guo}, Y.~Q. and {Guo}, Y.~Y. and {Han}, Y.~A. and {He}, H.~H. and {He}, H.~N. and {He}, J.~C. and {He}, S.~L. and {He}, X.~B. and {He}, Y. and {Heller}, M. and {Hor}, Y.~K. and {Hou}, C. and {Hou}, X. and {Hu}, H.~B. and {Hu}, S. and {Hu}, S.~C. and {Hu}, X.~J. and {Huang}, D.~H. and {Huang}, Q.~L. and {Huang}, W.~H. and {Huang}, X.~T. and {Huang}, Z.~C. and {Ji}, F. and {Ji}, X.~L. and {Jia}, H.~Y. and {Jiang}, K. and {Jiang}, Z.~J. and {Jin}, C. and {Kuleshov}, D. and {Levochkin}, K. and {Li}, B.~B. and {Li}, Cong and {Li}, Cheng and {Li}, F. and {Li}, H.~B. and {Li}, H.~C. and {Li}, H.~Y. and {Li}, J. and {Li}, K. and {Li}, W.~L. and {Li}, X. and {Li}, Xin and {Li}, X.~R. and {Li}, Y. and {Li}, Y.~Z. and {Li}, Zhe and {Li}, Zhuo and {Liang}, E.~W. and {Liang}, Y.~F. and {Lin}, S.~J. and {Liu}, B. and {Liu}, C. and {Liu}, D. and {Liu}, H. and {Liu}, H.~D. and {Liu}, J. and {Liu}, J.~L. and {Liu}, J.~S. and {Liu}, J.~Y. and {Liu}, M.~Y. and {Liu}, R.~Y. and {Liu}, S.~M. and {Liu}, W. and {Liu}, Y.~N. and {Liu}, Z.~X. and {Long}, W.~J. and {Lu}, R. and {Lv}, H.~K. and {Ma}, B.~Q. and {Ma}, L.~L. and {Ma}, X.~H. and {Mao}, J.~R. and {Masood}, A. and {Mitthumsiri}, W. and {Montaruli}, T. and {Nan}, Y.~C. and {Pang}, B.~Y. and {Pattarakijwanich}, P. and {Pei}, Z.~Y. and {Qi}, M.~Y. and {Ruffolo}, D. and {Rulev}, V. and {S{\'a}iz}, A. and {Shao}, L. and {Shchegolev}, O. and {Sheng}, X.~D. and {Shi}, J.~R. and {Song}, H.~C. and {Stenkin}, Yu. V. and {Stepanov}, V. and {Sun}, Q.~N. and {Sun}, X.~N. and {Sun}, Z.~B. and {Tam}, P.~H.~T. and {Tang}, Z.~B. and {Tian}, W.~W. and {Wang}, B.~D. and {Wang}, C. and {Wang}, H. and {Wang}, H.~G. and {Wang}, J.~C. and {Wang}, J.~S. and {Wang}, L.~P. and {Wang}, L.~Y. and {Wang}, R.~N. and {Wang}, W. and {Wang}, W. and {Wang}, X.~G. and {Wang}, X.~J. and {Wang}, X.~Y. and {Wang}, Y.~D. and {Wang}, Y.~J. and {Wang}, Y.~P. and {Wang}, Zheng and {Wang}, Zhen and {Wang}, Z.~H. and {Wang}, Z.~X. and {Wei}, D.~M. and {Wei}, J.~J. and {Wei}, Y.~J. and {Wen}, T. and {Wu}, C.~Y. and {Wu}, H.~R. and {Wu}, S. and {Wu}, W.~X. and {Wu}, X.~F. and {Xi}, S.~Q. and {Xia}, J. and {Xia}, J.~J. and {Xiang}, G.~M. and {Xiao}, G. and {Xiao}, H.~B. and {Xin}, G.~G. and {Xin}, Y.~L. and {Xing}, Y. and {Xu}, D.~L. and {Xu}, R.~X. and {Xue}, L. and {Yan}, D.~H. and {Yang}, C.~W. and {Yang}, F.~F. and {Yang}, J.~Y. and {Yang}, L.~L. and {Yang}, M.~J. and {Yang}, R.~Z. and {Yang}, S.~B. and {Yao}, Y.~H. and {Yao}, Z.~G. and {Ye}, Y.~M. and {Yin}, L.~Q. and {Yin}, N. and {You}, X.~H. and {You}, Z.~Y. and {Yu}, Y.~H. and {Yuan}, Q. and {Zeng}, H.~D. and {Zeng}, T.~X. and {Zeng}, W. and {Zeng}, Z.~K. and {Zha}, M. and {Zhai}, X.~X. and {Zhang}, B.~B. and {Zhang}, H.~M. and {Zhang}, H.~Y. and {Zhang}, J.~L. and {Zhang}, J.~W. and {Zhang}, L. and {Zhang}, Li and {Zhang}, L.~X. and {Zhang}, P.~F. and {Zhang}, P.~P. and {Zhang}, R. and {Zhang}, S.~R. and {Zhang}, S.~S. and {Zhang}, X. and {Zhang}, X.~P. and {Zhang}, Yong and {Zhang}, Yi and {Zhang}, Y.~F. and {Zhang}, Y.~L. and {Zhao}, B. and {Zhao}, J. and {Zhao}, L. and {Zhao}, L.~Z. and {Zhao}, S.~P. and {Zheng}, F. and {Zheng}, Y. and {Zhou}, B. and {Zhou}, H. and {Zhou}, J.~N. and {Zhou}, P. and {Zhou}, R. and {Zhou}, X.~X. and {Zhu}, C.~G. and {Zhu}, F.~R. and {Zhu}, H. and {Zhu}, K.~J. and {Zuo}, X.},
        title = "{Ultrahigh-energy photons up to 1.4 petaelectronvolts from 12 {\ensuremath{\gamma}}-ray Galactic sources}",
      journal = {\nat},
         year = 2021,
        month = jun,
       volume = {594},
       number = {7861},
        pages = {33-36},
          doi = {10.1038/s41586-021-03498-z},
       adsurl = {https://ui.adsabs.harvard.edu/abs/2021Natur.594...33C}
}

@ARTICLE{kothes+01,
       author = {{Kothes}, Roland and {Uyaniker}, B{\"u}lent and {Pineault}, Serge},
        title = "{The Supernova Remnant G106.3+2.7 and Its Pulsar-Wind Nebula: Relics of Triggered Star Formation in a Complex Environment}",
      journal = {\apj},
         year = 2001,
        month = oct,
       volume = {560},
       number = {1},
        pages = {236-243},
          doi = {10.1086/322511},
archivePrefix = {arXiv},
       eprint = {astro-ph/0106270},
 primaryClass = {astro-ph},
       adsurl = {https://ui.adsabs.harvard.edu/abs/2001ApJ...560..236K}
}

@ARTICLE{ricean,
       author = {{Wardle}, J.~F.~C. and {Kronberg}, P.~P.},
        title = "{The linear polarization of quasi-stellar radio sources at 3.71 and 11.1 centimeters.}",
      journal = {\apj},
         year = 1974,
        month = jan,
       volume = {194},
          eid = {249},
        pages = {249},
          doi = {10.1086/153240},
       adsurl = {https://ui.adsabs.harvard.edu/abs/1974ApJ...194..249W}
}

@ARTICLE{kothes+08,
       author = {{Kothes}, R. and {Landecker}, T.~L. and {Reich}, W. and {Safi-Harb}, S. and {Arzoumanian}, Z.},
        title = "{DA 495: An Aging Pulsar Wind Nebula}",
      journal = {\apj},
         year = 2008,
        month = nov,
       volume = {687},
       number = {1},
        pages = {516-531},
          doi = {10.1086/591653},
archivePrefix = {arXiv},
       eprint = {0807.0811},
 primaryClass = {astro-ph},
       adsurl = {https://ui.adsabs.harvard.edu/abs/2008ApJ...687..516K}
}

@ARTICLE{conto+99,
       author = {{Contopoulos}, Ioannis and {Kazanas}, Demosthenes and {Fendt}, Christian},
        title = "{The Axisymmetric Pulsar Magnetosphere}",
      journal = {\apj},
         year = 1999,
        month = jan,
       volume = {511},
       number = {1},
        pages = {351-358},
          doi = {10.1086/306652},
archivePrefix = {arXiv},
       eprint = {astro-ph/9903049},
 primaryClass = {astro-ph},
       adsurl = {https://ui.adsabs.harvard.edu/abs/1999ApJ...511..351C}
}

@ARTICLE{Chevalier+11,
       author = {{Chevalier}, Roger A. and {Reynolds}, Stephen P.},
        title = "{Pulsar Wind Nebulae with Thick Toroidal Structure}",
      journal = {\apjl},
         year = 2011,
        month = oct,
       volume = {740},
       number = {1},
          eid = {L26},
        pages = {L26},
          doi = {10.1088/2041-8205/740/1/L26},
archivePrefix = {arXiv},
       eprint = {1109.1752},
 primaryClass = {astro-ph.HE},
       adsurl = {https://ui.adsabs.harvard.edu/abs/2011ApJ...740L..26C}
}

@ARTICLE{haverkorn+08,
       author = {{Haverkorn}, M. and {Brown}, J.~C. and {Gaensler}, B.~M. and {McClure-Griffiths}, N.~M.},
        title = "{The Outer Scale of Turbulence in the Magnetoionized Galactic Interstellar Medium}",
      journal = {\apj},
         year = 2008,
        month = jun,
       volume = {680},
       number = {1},
        pages = {362-370},
          doi = {10.1086/587165},
archivePrefix = {arXiv},
       eprint = {0802.2740},
 primaryClass = {astro-ph},
       adsurl = {https://ui.adsabs.harvard.edu/abs/2008ApJ...680..362H}
}

@ARTICLE{minter+96,
       author = {{Minter}, Anthony H. and {Spangler}, Steven R.},
        title = "{Observation of Turbulent Fluctuations in the Interstellar Plasma Density and Magnetic Field on Spatial Scales of 0.01 to 100 Parsecs}",
      journal = {\apj},
         year = 1996,
        month = feb,
       volume = {458},
        pages = {194},
          doi = {10.1086/176803},
       adsurl = {https://ui.adsabs.harvard.edu/abs/1996ApJ...458..194M}
}

@ARTICLE{ng04,
       author = {{Ng}, C. -Y. and {Romani}, Roger W.},
        title = "{Fitting Pulsar Wind Tori}",
      journal = {\apj},
         year = 2004,
        month = jan,
       volume = {601},
       number = {1},
        pages = {479-484},
          doi = {10.1086/380486},
archivePrefix = {arXiv},
       eprint = {astro-ph/0310155},
 primaryClass = {astro-ph},
       adsurl = {https://ui.adsabs.harvard.edu/abs/2004ApJ...601..479N}
}

@ARTICLE{weisskopf+00,
       author = {{Weisskopf}, Martin C. and {Hester}, J. Jeff and {Tennant}, Allyn F. and {Elsner}, Ronald F. and {Schulz}, Norbert S. and {Marshall}, Herman L. and {Karovska}, Margarita and {Nichols}, Joy S. and {Swartz}, Douglas A. and {Kolodziejczak}, Jeffery J. and {O'Dell}, Stephen L.},
        title = "{Discovery of Spatial and Spectral Structure in the X-Ray Emission from the Crab Nebula}",
      journal = {\apjl},
         year = 2000,
        month = jun,
       volume = {536},
       number = {2},
        pages = {L81-L84},
          doi = {10.1086/312733},
archivePrefix = {arXiv},
       eprint = {astro-ph/0003216},
 primaryClass = {astro-ph},
       adsurl = {https://ui.adsabs.harvard.edu/abs/2000ApJ...536L..81W}
}

@ARTICLE{burn66,
       author = {{Burn}, B.~J.},
        title = "{On the depolarization of discrete radio sources by Faraday dispersion}",
      journal = {\mnras},
         year = 1966,
        month = jan,
       volume = {133},
        pages = {67},
          doi = {10.1093/mnras/133.1.67},
       adsurl = {https://ui.adsabs.harvard.edu/abs/1966MNRAS.133...67B}
}

@ARTICLE{jones_odell77,
       author = {{Jones}, T.~W. and {Odell}, S.~L.},
        title = "{Transfer of polarized radiation in self-absorbed synchrotron sources. I. Results for a homogeneous source.}",
      journal = {\apj},
         year = 1977,
        month = jun,
       volume = {214},
        pages = {522-539},
          doi = {10.1086/155278},
       adsurl = {https://ui.adsabs.harvard.edu/abs/1977ApJ...214..522J}
}

@ARTICLE{snail,
       author = {{Ma}, Y.~K. and {Ng}, C. -Y. and {Bucciantini}, N. and {Slane}, P.~O. and {Gaensler}, B.~M. and {Temim}, T.},
        title = "{Radio Polarization Observations of the Snail: A Crushed Pulsar Wind Nebula in G327.1-1.1 with a Highly Ordered Magnetic Field}",
      journal = {\apj},
         year = 2016,
        month = apr,
       volume = {820},
       number = {2},
          eid = {100},
        pages = {100},
          doi = {10.3847/0004-637X/820/2/100},
archivePrefix = {arXiv},
       eprint = {1603.04886},
 primaryClass = {astro-ph.HE},
       adsurl = {https://ui.adsabs.harvard.edu/abs/2016ApJ...820..100M}
}

@BOOK{pac70,
       author = {{Pacholczyk}, A.~G.},
        title = "{Radio astrophysics. Nonthermal Processes in Galactic and Extragalactic Sources}",
         year = 1970,
       adsurl = {https://ui.adsabs.harvard.edu/abs/1970ranp.book.....P},
      publisher = {Freeman},
      address = {San Francisco, CA}
}

@ARTICLE{olmi+16,
       author = {{Olmi}, B. and {Del Zanna}, L. and {Amato}, E. and {Bucciantini}, N. and {Mignone}, A.},
        title = "{Multi-D magnetohydrodynamic modelling of pulsar wind nebulae: recent progress and open questions}",
      journal = {Journal of Plasma Physics},
         year = 2016,
        month = dec,
       volume = {82},
       number = {6},
          eid = {635820601},
        pages = {635820601},
          doi = {10.1017/S0022377816000957},
archivePrefix = {arXiv},
       eprint = {1610.07956},
 primaryClass = {astro-ph.HE},
       adsurl = {https://ui.adsabs.harvard.edu/abs/2016JPlPh..82f6301O}
}

@ARTICLE{porth+14,
       author = {{Porth}, Oliver and {Komissarov}, Serguei S. and {Keppens}, Rony},
        title = "{Three-dimensional magnetohydrodynamic simulations of the Crab nebula}",
      journal = {\mnras},
         year = 2014,
        month = feb,
       volume = {438},
       number = {1},
        pages = {278-306},
          doi = {10.1093/mnras/stt2176},
archivePrefix = {arXiv},
       eprint = {1310.2531},
 primaryClass = {astro-ph.HE},
       adsurl = {https://ui.adsabs.harvard.edu/abs/2014MNRAS.438..278P}
}

@ARTICLE{gaensler+wallace03,
       author = {{Gaensler}, B.~M. and {Wallace}, B.~J.},
        title = "{A Multifrequency Radio Study of Supernova Remnant G292.0+1.8 and Its Pulsar Wind Nebula}",
      journal = {\apj},
         year = 2003,
        month = sep,
       volume = {594},
       number = {1},
        pages = {326-339},
          doi = {10.1086/376861},
archivePrefix = {arXiv},
       eprint = {astro-ph/0305168},
 primaryClass = {astro-ph},
       adsurl = {https://ui.adsabs.harvard.edu/abs/2003ApJ...594..326G}
}

@ARTICLE{govoni+04,
       author = {{Govoni}, Federica and {Feretti}, Luigina},
        title = "{Magnetic Fields in Clusters of Galaxies}",
      journal = {International Journal of Modern Physics D},
         year = 2004,
        month = jan,
       volume = {13},
       number = {8},
        pages = {1549-1594},
          doi = {10.1142/S0218271804005080},
archivePrefix = {arXiv},
       eprint = {astro-ph/0410182},
 primaryClass = {astro-ph},
       adsurl = {https://ui.adsabs.harvard.edu/abs/2004IJMPD..13.1549G}
}

@ARTICLE{kc84,
       author = {{Kennel}, C.~F. and {Coroniti}, F.~V.},
        title = "{Confinement of the Crab pulsar's wind by its supernova remnant.}",
      journal = {\apj},
         year = 1984,
        month = aug,
       volume = {283},
        pages = {694-709},
          doi = {10.1086/162356},
       adsurl = {https://ui.adsabs.harvard.edu/abs/1984ApJ...283..694K}
}

@ARTICLE{crab_mag,
       author = {{Bietenholz}, M.~F. and {Kronberg}, P.~P.},
        title = "{The Magnetic Field of the Crab Nebula and the Nature of Its ``Jet''}",
      journal = {\apjl},
         year = 1990,
        month = jul,
       volume = {357},
        pages = {L13},
          doi = {10.1086/185755},
       adsurl = {https://ui.adsabs.harvard.edu/abs/1990ApJ...357L..13B}
}

@ARTICLE{kothes06,
       author = {{Kothes}, Roland and {Reich}, Wolfgang and {Uyan{\i}ker}, B{\"u}lent},
        title = "{The Boomerang PWN G106.6+2.9 and the Magnetic Field Structure in Pulsar Wind Nebulae}",
      journal = {\apj},
         year = 2006,
        month = feb,
       volume = {638},
       number = {1},
        pages = {225-233},
          doi = {10.1086/498666},
       adsurl = {https://ui.adsabs.harvard.edu/abs/2006ApJ...638..225K}
}

@ARTICLE{dlmd03,
       author = {{Dodson}, R. and {Lewis}, D. and {McConnell}, D. and {Deshpande}, A.~A.},
        title = "{The radio nebula surrounding the Vela pulsar}",
      journal = {\mnras},
         year = 2003,
        month = jul,
       volume = {343},
       number = {1},
        pages = {116-124},
          doi = {10.1046/j.1365-8711.2003.06653.x},
archivePrefix = {arXiv},
       eprint = {astro-ph/0302373},
 primaryClass = {astro-ph},
       adsurl = {https://ui.adsabs.harvard.edu/abs/2003MNRAS.343..116D}
}

@ARTICLE{ng08,
       author = {{Ng}, C. -Y. and {Romani}, Roger W.},
        title = "{Fitting Pulsar Wind Tori. II. Error Analysis and Applications}",
      journal = {\apj},
         year = 2008,
        month = jan,
       volume = {673},
       number = {1},
        pages = {411-417},
          doi = {10.1086/523935},
archivePrefix = {arXiv},
       eprint = {0710.4168},
 primaryClass = {astro-ph},
       adsurl = {https://ui.adsabs.harvard.edu/abs/2008ApJ...673..411N}
}

@INPROCEEDINGS{reich02,
       author = {{Reich}, W.},
        title = "{Radio Observations of Supernova Remnants}",
    booktitle = {Neutron Stars, Pulsars, and Supernova Remnants},
         year = 2002,
       editor = {{Becker}, W. and {Lesch}, H. and {Tr{\"u}mper}, J.},
        month = jan,
        pages = {1},
archivePrefix = {arXiv},
       eprint = {astro-ph/0208498},
 primaryClass = {astro-ph},
       adsurl = {https://ui.adsabs.harvard.edu/abs/2002nsps.conf....1R}
}

@ARTICLE{casa,
       author = {{CASA Team} and {Bean}, Ben and {Bhatnagar}, Sanjay and {Castro}, Sandra and {Donovan Meyer}, Jennifer and {Emonts}, Bjorn and {Garcia}, Enrique and {Garwood}, Robert and {Golap}, Kumar and {Gonzalez Villalba}, Justo and {Harris}, Pamela and {Hayashi}, Yohei and {Hoskins}, Josh and {Hsieh}, Mingyu and {Jagannathan}, Preshanth and {Kawasaki}, Wataru and {Keimpema}, Aard and {Kettenis}, Mark and {Lopez}, Jorge and {Marvil}, Joshua and {Masters}, Joseph and {McNichols}, Andrew and {Mehringer}, David and {Miel}, Renaud and {Moellenbrock}, George and {Montesino}, Federico and {Nakazato}, Takeshi and {Ott}, Juergen and {Petry}, Dirk and {Pokorny}, Martin and {Raba}, Ryan and {Rau}, Urvashi and {Schiebel}, Darrell and {Schweighart}, Neal and {Sekhar}, Srikrishna and {Shimada}, Kazuhiko and {Small}, Des and {Steeb}, Jan-Willem and {Sugimoto}, Kanako and {Suoranta}, Ville and {Tsutsumi}, Takahiro and {van Bemmel}, Ilse M. and {Verkouter}, Marjolein and {Wells}, Akeem and {Xiong}, Wei and {Szomoru}, Arpad and {Griffith}, Morgan and {Glendenning}, Brian and {Kern}, Jeff},
        title = "{CASA, the Common Astronomy Software Applications for Radio Astronomy}",
      journal = {\pasp},
         year = 2022,
        month = nov,
       volume = {134},
       number = {1041},
          eid = {114501},
        pages = {114501},
          doi = {10.1088/1538-3873/ac9642},
archivePrefix = {arXiv},
       eprint = {2210.02276},
 primaryClass = {astro-ph.IM},
       adsurl = {https://ui.adsabs.harvard.edu/abs/2022PASP..134k4501C}
}
\bibliographystyle{aasjournalv7}

\end{document}